\documentclass[twocolumn]{revtex4}
\usepackage{amsmath}
\usepackage{amssymb}
\usepackage{rsfs}
\usepackage{color}
\usepackage{ulem}
\newcommand{\rra}{\rangle\!\rangle}
\newcommand{\lla}{\langle\!\langle}
\newcommand{\ra}{\rangle}
\newcommand{\la}{\langle}

\begin{document}

\title{Euclidean and Lorentzian Actions of the Classicalized Holographic Tensor Network}

\author{Eiji Konishi}
\email{konishi.eiji.27c@kyoto-u.jp}
\address{Graduate School of Human and Environmental Studies, Kyoto University, Kyoto 606-8501, Japan}

\date{\today}

\begin{abstract}
In three spacetime dimensions, we propose a generally covariant Lorentzian action of the classicalized holographic tensor network (cHTN) as the holographic reduction of the Einstein--Hilbert action of gravity in the presence of a negative cosmological constant.
In this article, first, we investigate the properties of this Lorentzian action in the ground state.
Next, based on the Euclidean action of the cHTN, we derive the gravity perturbation induced by a massive particle at rest in the cHTN as the Unruh effect.
Finally, we view our holographic formulation of spacetime as a non-equilibrium second law subject to general covariance.
\end{abstract}

\maketitle

\section{Introduction}

In the conventional interpretation of quantum gravity \cite{Kuchar,Isham} without the holographic principle \cite{Hol1,Hol2,Hol3}, the quantum state is that of the whole Universe.
A typical application of the Born rule in this interpretation is seen in the inflationary multiverse scenario \cite{CC1,CC2,Vilenkin}.

Taking a different approach, the author has proposed a novel interpretation of quantum gravity \cite{EPL1,EPL2} based on the holographic principle \cite{Hol1,Hol2,Hol3} in the context of the three-dimensional anti-de Sitter spacetime/two-dimensional conformal field theory (AdS$_3$/CFT$_2$) correspondence \cite{AdSCFT1,AdSCFT2,AdSCFT3,AdSCFT4} at the strong-coupling limit of the boundary CFT$_2$ \cite{RT1,RT2,HRT,RT3,Swingle,Matsueda,Review1,Review2,Review3}.
In this interpretation of quantum gravity, non-selective quantum measurement \cite{dEspagnat} of the ground state or a purified quantum thermal equilibrium state of space, that is, a holographic tensor network (HTN) \cite{Swingle,Matsueda,Review1}, is done in the ensemble interpretation of quantum mechanics by decohering quantum coherence in this quantum state completely.
The decoherence (i.e., loss of quantum interference with respect to the observables) is exactly done by introducing a superselection rule operator and then restricting the set of observables acting on the Hilbert space of the HTN to the Abelian set whose elements commute with the superselection rule operator \cite{JHAP1}.
The author refers to this decoherence as {\it classicalization}.
The classicalization of quantum gravity is {\it not} classical gravity; indeed, the classicalized state of the HTN is still a quantum state but a highly non-trivial mixed state.
Since this quantum state is a statistical mixture of the product quantum eigenstates, there are {\it negative} local degrees of freedom \cite{EPL2,JHAP1}.

So far, we have classicalized space in the Euclidean regime of the HTN, that is, the purified quantum thermal equilibrium states of the boundary CFT$_2$ including the ground state \cite{EPL1,EPL2,JHAP1,JHAP2}.
Then, to formulate the time-dependent HTN in the Lorentzian regime, how do we incorporate real time $t$ into this interpretation of quantum gravity?
The answer proposed in this article is to {\it classicalize real time, too} \cite{JPCO}.
Namely, we treat real time as a classical observable {\it $\grave{a}$ la} von Neumann \cite{dEspagnat,Neumann} in the exact sense.
Here, we treat real time continuously.
Then, the Hilbert space of the HTN can be decomposed into a direct integral of the continuous coherent subspaces with an absolutely continuous temporal measure $d\mu(t)$ of the density matrix \cite{AnnMath}.

In this article, we propose a generally covariant Lorentzian action of the classicalized holographic tensor network (cHTN) as the holographic reduction of the Einstein--Hilbert action of gravity in three spacetime dimensions
\begin{equation}
I_{\rm EH}[g_{\mu\nu}]=\frac{1}{16 \pi G_N}\int (R-2\Lambda)\sqrt{-g}d^2x dt\label{eq:EH}
\end{equation}
in the presence of a negative cosmological constant $\Lambda$.
Here, we choose $(-,+,+)$ as the signature of the Lorentzian spacetime metric $g_{\mu\nu}$, and $G_N$ and $R$ denote the three-dimensional Newtonian gravitational constant and the scalar curvature of the Lorentzian spacetime metric $g_{\mu\nu}$, respectively.
Our Lorentzian action of the cHTN is defined for a generic quantum pure state $|\psi(t)\ra$ of the boundary CFT$_2$ by
\begin{equation}
I_L[|\psi\rra_L]=-\hbar H[|\psi\rra_L]\;,\label{eq:H}
\end{equation}
where we define the quantum state of the HTN in the representation of the Lorentzian boundary conformal symmetry
\begin{equation}
|\psi\rra_L\equiv\int^\bigoplus|\psi(t)\ra \sqrt{d\mu(t)}\;.\label{eq:rra}
\end{equation}
$H[|\psi\rra_L]$ is the von Neumann entropy (here, the measurement entropy) of the classicalized state of $|\psi\rra_L$ in nats.
In addition to the Euclidean action \cite{EPL2}, this Lorentzian action accords with the holographic principle \cite{Hol1,Hol2,Hol3} and asserts that the negative number of the local {\it spin} degrees of freedom in the bulk spacetime is given by the amount of boundary CFT$_2$ state information in nats that is lost by the classicalization \cite{EPL2}.
Note that, if $|\psi\rra_L$ is the ground state, it is time independent (i.e., a pure state with respect to real time), and thus $|\psi\rra_L {}_L\lla \psi|=|\psi\ra\la \psi|\otimes \widehat{1}$ and $H[|\psi\rra_L]=H[|\psi\ra]$ hold.
In this case, our Lorentzian action (\ref{eq:H}) of the cHTN is reduced to the Euclidean action of the cHTN
\begin{equation}
I_E[|\psi\ra]=-\hbar H[|\psi\ra]\;,\label{eq:H20}
\end{equation}
which was used by the author in Refs. \cite{EPL2,JHAP1,JHAP2}.
In the Euclidean action (\ref{eq:H20}) of the cHTN, $|\psi\ra$ is the ground state or a purified quantum thermal equilibrium state of the HTN \cite{EPL2}.
Subject to a given average energy, the HTN in the Euclidean regime is the most probable statistical mixture with respect to {\it energy}, and the HTN in the Lorentzian regime is now the most probable statistical mixture with respect to {\it real time}.
Here, in the HTN, there is a quantum uncertainty relation between energy and real time because energy and real time are conjugate to each other.

However, after the classicalization of the HTN, there is no quantum uncertainty relation between them.
So, in the presence or absence of matter, the cHTN in the Lorentzian regime is the most probable statistical mixture with respect to energy {\it and} real time {\it simultaneously}.
Based on this fact, we introduce the imaginary time $\tau \equiv it$ and extend the Euclidean action of the cHTN from Eq. (\ref{eq:H20}) to
\begin{equation}
I_E[|\psi\rra_E]=-\hbar H[|\psi\rra_E]\;,\label{eq:H2}
\end{equation}
where we define the quantum state of the HTN in the representation of the Euclidean boundary conformal symmetry
\begin{equation}
|\psi\rra_E\equiv \int^\bigoplus |\psi(\tau)\ra \sqrt{d\mu (\tau)}\label{eq:rra2}
\end{equation}
for an absolutely continuous temporal measure $d\mu(\tau)$ of the density matrix, and $H[|\psi\rra_E]$ is the measurement entropy of the classicalized state of $|\psi\rra_E$ in nats.
Here, this imaginary time $\tau$ is the real-valued time coordinate in the Euclidean spacetime and is distinguished from the inverse temperature of the quantum thermal equilibrium states of the HTN, which is the Lagrange multiplier for the fixed average energy of the HTN, except for the identification of the period of the imaginary time with the inverse temperature \cite{GH,Harlow}.\footnote{For the issue of allowing complex-valued spacetime metrics in Euclidean quantum gravity, see Ref. \cite{Witten} and the references therein.}
Then, in the presence or absence of matter, the cHTN in the Euclidean regime is the most probable statistical mixture with respect to energy {\it and} imaginary time {\it simultaneously}.

From the results in Ref. \cite{JHAP1}, in the cHTN of the HTN in the ground state, the Euclidean regime is more fundamental than the Lorentzian regime because bulk quantum mechanics of a non-relativistic free particle in the Lorentzian regime follows from the bulk classical stochastic process of this particle (i.e., the readout process of local spin events by the classicalized hologram) in the Euclidean regime via the inverse Wick rotation
\begin{equation}
t=-i \tau\;.\label{eq:Wick}
\end{equation}
Here, in the Lorentzian regime, quantum measuring systems with the ability to read out events \cite{EPL3} exist only in the bulk spacetime; in the Euclidean regime, on the other hand, the classicalized hologram on the boundary spacetime is the only quantum measuring system.

In the rest of this article, we study the properties of the Euclidean and Lorentzian actions of the cHTN.
In Sec. II, we investigate the properties of the proposed Lorentzian action (\ref{eq:H}) of the cHTN in the ground state.
In Sec. III, we derive the gravity perturbation induced by a massive particle at rest in the cHTN from the Euclidean action (\ref{eq:H2}) of the cHTN as the Unruh effect \cite{Davies,Unruh1,Sewell,Unruh2,Harlow2}.
In Sec. IV, we conclude the article by arguing that our holographic formulation of spacetime can be viewed as a non-equilibrium second law subject to general covariance.

\section{Ground-state properties of the Lorentzian action}

In this section, we show three properties of the Lorentzian action (\ref{eq:H}) of the cHTN in the ground state:
\begin{enumerate}
\item[(I)] In the ground state, the Lorentzian action (\ref{eq:H}) of the cHTN is the holographic reduction of the Einstein--Hilbert action (\ref{eq:EH}).

\item[(II)] In the ground state, the proposed Lorentzian action (\ref{eq:H}) of the cHTN is generally covariant.

\item[(III)] The Lorentzian AdS$_3$ spacetime metric can be recovered from the ground state of the boundary CFT$_2$.
\end{enumerate}

Here, the {\it ground state} refers to that of the Einstein--Hilbert action (\ref{eq:EH}) and that of the Lorentzian action (\ref{eq:H}) of the cHTN when there are no additional actions (i.e., there is gravity and a negative cosmological constant only).

\subsection{Holographic reduction}

We consider the ground state, which is a pure state with respect to real time, in the Hilbert space of the boundary CFT$_2$.
The quantum entanglement folded in the boundary ground state is unfolded to the multi-scale entanglement renormalization ansatz (MERA) of this state \cite{Vidal1,Vidal2} along the extra spatial dimension in the bulk space \cite{Swingle}.
We unfold the Lorentzian action (\ref{eq:H}) of the cHTN also into the bulk space from the boundary.
Then, the measurement entropy $H[|\psi\rra_L]$ of the cHTN in bits is given by the discretized area of the MERA \cite{EPL1,EPL2}.
Because the pixel of the MERA is given by $R_{\rm AdS}^2$ for the curvature radius $R_{\rm AdS}$ of the AdS$_3$, the on-shell local information density, $\eta$, is given by
\begin{equation}
\eta=-\frac{1}{R_{\rm AdS}^2}\;.\label{eq:eta}
\end{equation}
Here, the number $1$ represents the spatial dimensions of the boundary.
Note that, in the flat spacetime limit, $\eta$ converges to zero.

On the other hand, the on-shell solution, that is, the Lorentzian AdS$_3$ spacetime of the Einstein--Hilbert action (\ref{eq:EH}), is a maximally symmetric spacetime and has a negative constant scalar curvature.
Because its Ricci tensor is $R_{\mu\nu}=(1-3)g_{\mu\nu}/R_{\rm AdS}^2$ \cite{Nastase}, the on-shell local information density of the gravity part of the Einstein--Hilbert action (\ref{eq:EH}) is given by
\begin{equation}
\eta_{\rm EH}=\frac{R}{2}=-\frac{3}{R_{\rm AdS}^2}\;.\label{eq:eta2}
\end{equation}
Here, the number $3$ represents half the number of off-diagonal elements (i.e., the number of plane combinations) in a square matrix of order $3$ (i.e., the spacetime dimensions).
From this and Eq. (\ref{eq:eta}), in the ground state, the Lorentzian action (\ref{eq:H}) of the cHTN is the holographic reduction of the Einstein--Hilbert action (\ref{eq:EH}).

\subsection{General covariance}

First, the inner product ${}_L\lla \psi_2|\psi_1\rra_L$ between two generic states $|\psi_1\rra_L$ and $|\psi_2\rra_L$ of Eq. (\ref{eq:rra}) is invariant under the change of the temporal measure $d\mu(t)$ to another temporal measure $d\nu(t)$ used in the direct integral decomposition to which the same temporal resolution of unity $\widehat{1}$ belongs \cite{AnnMath}.
This means that the unitary equivalence class of the Hilbert space of $|\psi\rra_L$ and thus the measurement entropy $H[|\psi\rra_L]$ are independent from the choice of temporal measure in the direct integral decomposition, and thus the Lorentzian action (\ref{eq:H}) of the cHTN is well-defined for a generic state $|\psi\rra_L$ of Eq. (\ref{eq:rra}).

Next, in the ground state, from Eqs. (\ref{eq:eta}) and (\ref{eq:eta2}), the Lorentzian action (\ref{eq:H}) of the cHTN has one locally independent negative degree of freedom per pixel $R_{\rm AdS}^2$, the same as the Einstein--Hilbert action (\ref{eq:EH}), to gauge the symmetry spatial coordinate transformations on the boundary CFT$_2$ to the spatial diffeomorphisms in the bulk.
In the cHTN, the bulk spatial diffeomorphisms are unitary transformations (i.e., classical gauge transformations) on the diagonal classicalized state of $|\psi(t)\ra$, and these are enhanced to the bulk spacetime diffeomorphisms as unitary transformations on the diagonal classicalized state of $|\psi\rra_L$.
Then, the Lorentzian action (\ref{eq:H}) of the cHTN is invariant under the bulk spacetime diffeomorphisms (i.e., generally covariant) because the von Neumann entropy is invariant under the unitary transformation.

\subsection{Recovery of the Lorentzian spacetime metric}

We recover the Lorentzian AdS$_3$ spacetime metric $g_{\mu\nu}$ from the ground state of the boundary CFT$_2$.\footnote{In this subsection, we set $R_{\rm AdS}=1$.}
In Ref. \cite{RINP}, after averaging with respect to the local spin degree of freedom over the statistical mixture of its two eigenstates at each site of the cHTN, we recovered the spatial metric of a real-time slice of the Lorentzian AdS$_3$ spacetime
\begin{equation}
ds^2|_{dt=0}=\frac{dx^2+dr^2}{r^2}\;,\label{eq:metric1}
\end{equation}
where $x$ and $r$ are the rescaled horizontal and redefined radial coordinates of the sites of the MERA, respectively.
Now, we regard $x$ and $r$ as {\it spatial} coordinates and incorporate real time $t$ into this previous result.
Due to the conformal $SO(2,2)$ symmetry of the Lorentzian boundary CFT$_2$, the Lorentzian bulk spacetime has the $SO(2,2)$ isometry group.
From this spacetime symmetry and Eq. (\ref{eq:metric1}), we obtain the static Lorentzian spacetime metric
\begin{equation}
ds^2=-\frac{f(r)dt^2}{r^2}+ds^2|_{dt=0}\label{eq:metric2}
\end{equation}
for a dimensionless function $f(r)$.
Here, note that the ground state is the thermal equilibrium state at zero temperature and thus has no length-scale variable in natural units.
As a result of this fact and Eq. (\ref{eq:metric2}), we recover the Lorentzian AdS$_3$ spacetime metric
\begin{equation}
ds^2=\frac{-dt^2+dx^2+dr^2}{r^2}\;,\label{eq:metric3}
\end{equation}
where the coordinates $t$ and $x$ in Eq. (\ref{eq:metric3}) on the $r=0$ slice without the conformal factor match those of the boundary spacetime \cite{Nastase}.

\section{Gravity as the Unruh effect}

In this section, in the Euclidean regime of the HTN, we derive the gravity perturbation that is the Wick-rotated proper {\it acceleration} induced by a massive particle at rest in the cHTN as the Unruh effect.

We assume a particle with non-zero mass $M$ located at the {\it top tensor} of the cHTN \cite{Vidal2} and study its effect over the cHTN.
After an infinitesimal imaginary proper time interval $d \tau_0$ at the top tensor, this mass $M$ of the particle creates an infinitesimal spin-information reading in nats with an amount
\begin{equation}
dI_0=\frac{d_\tau S_E}{\hbar}\;,\ \ d_\tau S_E=Mc^2 d \tau_0\;,
\end{equation}
where
\begin{equation}
S_E\equiv -iS_L|_{t\to -i\tau}
\end{equation}
is the Wick rotation of the relativistic action $S_L$ of the particle $M$ \cite{WickCQG} and is added to the Euclidean action (\ref{eq:H2}) of the cHTN.
This infinitesimal spin-information reading $dI_0$ at the top tensor would be fine-grained in the cHTN along the inverse renormalization group (RG) direction of the ground state of the boundary CFT$_2$ and equally divided per site at each deeper inverse RG step $n$ counted from the top tensor.
Then, at the inverse RG step $n$, the infinitesimal spin-information reading $dI_0$ is fine-grained to a smaller amount of spin-information reading per site
\begin{equation}
di_{0\to n}=\frac{dI_0}{N_n}\;,
\end{equation}
where $N_n$ is the number of sites in the cHTN at the inverse RG step $n$.
This smaller amount of infinitesimal spin-information reading $di_{0 \to n}$ per site is equivalent to a finite energy per site
\begin{equation}
\varepsilon_n=\hbar \frac{di_{0\to n}}{d \tau_n}
\end{equation}
for the infinitesimal imaginary proper time interval $d \tau_n$ at the inverse RG step $n$.
Now, we note two facts: there is local von Neumann entropy $\sigma$ of $1$ nat at every site of the cHTN \cite{RINP}, and the cHTN is originally in the ground state.
Because of these two facts, per site, this absent energy $\varepsilon_n$ defines the physical Unruh temperature $T^U_n$ (see remark (i)) by \cite{Sagawa}
\begin{equation}
\sigma k_BT^U_n \equiv \varepsilon_n\;,\ \ \sigma=1\;.\label{eq:defTn}
\end{equation}
As the Unruh effect \cite{Davies,Unruh1,Sewell,Unruh2,Harlow2}, this physical Unruh temperature $T_n^U$, with the boost generator as the Hamiltonian \cite{Unruh2}, is created by the physical Wick-rotated proper acceleration, of magnitude $a_n$, of the observational frame of reference as
\begin{eqnarray}
\frac{\hbar a_n}{2\pi c}=k_BT^U_n=\frac{Mc^2}{N_n}\frac{d \tau_0}{d \tau_n}\;.\label{eq:Unruh}
\end{eqnarray}
From this, we arrive at the final formula
\begin{equation}
a_n=\frac{2\pi c^3}{\hbar}\frac{M}{N_n}\frac{d \tau_0}{d \tau_n}\;,\label{eq:an}
\end{equation}
where the direction of the Wick-rotated proper acceleration maximally increases the coarse grain of the infinitesimal spin-information reading $dI_0$ toward the top tensor, where the particle $M$ is located, and matches the forward RG direction of the ground state of the boundary CFT$_2$.
Now, $a_n$ is the Wick-rotated proper acceleration, induced by the particle $M$ located at the top tensor, in the cHTN at the inverse RG step $n$.
Note that, in the context of general relativity, the Lorentzian gravitational proper acceleration in the cHTN at the inverse RG step $n$ is identically zero, since gravity is not a real force but a curved spacetime.
However, since we fix the Lorentzian spacetime metric $g_{\mu\nu}$ to the background static Lorentzian spacetime metric (\ref{eq:metric3}) recovered from the ground state of the boundary CFT$_2$, we can interpret this Wick-rotated proper acceleration $a_n$ as a gravity perturbation in the cHTN in the Euclidean regime of the HTN.
Next, the Wick-rotated proper acceleration, induced by a particle with non-zero mass $m$ located at a site in the cHTN at the inverse RG step $n$, in the cHTN at the top tensor is given by
\begin{equation}
a_0=\frac{2\pi c^3}{\hbar}\frac{m}{N_0}\frac{d \tau_n}{d \tau_0}\;.\label{eq:an2}
\end{equation}
Here, the direction of the Wick-rotated proper acceleration maximally increases the coarse grain (i.e., maximally decreases the inverse RG step $n$) of the infinitesimal spin-information reading $dI_n$ toward the site where the particle $m$ is located; $dI_n$ is created by the particle $m$ after the infinitesimal imaginary proper time interval $d \tau_n$ and is coarse-grained to $di_{n\to 0}=dI_n/N_0$ at the top tensor.
Then, from Eqs. (\ref{eq:an}) and (\ref{eq:an2}), the consistency of this interpretation, that is, the conservation of Wick-rotated three-momentum, $p^\mu$, in the system of the particle $M$ with $p_0^\mu$ and the particle $m$ with $p_n^\mu$, holds as
\begin{equation}
p_0^r\sqrt{\gamma|_0}+p_n^r\sqrt{\gamma|_n}=0
\end{equation}
for the Wick-rotated metric $\gamma_{\mu\nu}|_n$ of the Euclidean spacetime at the inverse RG step $n$.\footnote{Here, $a_0^r=-a_0/\sqrt{\gamma_{rr}|_0}$ and $a_n^r=a_n/\sqrt{\gamma_{rr}|_n}$ hold.}

We make three remarks:
\begin{enumerate}
\item[(i)] The Unruh temperature $T^U_n$ defined by Eq. (\ref{eq:defTn}) is {\it physical} in the sense that it is not created by a gauge.

\item[(ii)] In Eq. (\ref{eq:Unruh}), $\pi c/a_n$ is the half period of the Wick-rotated orbit of the boost Killing field, specified by $a_n$ \cite{Davies,Harlow2}.

\item[(iii)] In Eq. (\ref{eq:Unruh}), the Fock space is defined for a spatiotemporally local patch and thus differs from the Fock space which is used in Refs. \cite{DL1,DL2}.
\end{enumerate}

In this derivation of the gravity perturbation induced by a massive particle located at the top tensor of the cHTN from the Euclidean action (\ref{eq:H2}) of the cHTN, three points are essential:
\begin{enumerate}
\item[(A)] The Wick-rotated relativistic action of a massive particle at rest generates spin-information reading after an imaginary time interval.

\item[(B)] Spin-information reading at the top tensor would be fine-grained in the cHTN along the inverse RG direction of the ground state of the boundary CFT$_2$.

\item[(C)] There is local von Neumann entropy of $1$ nat at every site of the cHTN \cite{RINP}.
\end{enumerate}
Because of this, the physical Unruh temperature $T^U_n$ is defined from the rest energy $Mc^2$ of the particle.
Then, in the Euclidean regime of the HTN, a gravity perturbation is induced by the particle $M$ in the cHTN as the Unruh effect.

\section{Conclusion}

In this article, we studied three subjects.
First, we proposed the generally covariant Lorentzian action of the cHTN by classicalizing real time in addition to the HTN.
Second, we investigated the properties of this Lorentzian action of the cHTN in the ground state.
Third, based on the Euclidean action of the cHTN, we derived the gravity perturbation induced by a massive particle at rest in the cHTN as the Unruh effect.

Our Euclidean and Lorentzian actions of the cHTN do not require the minimum action principle of classical mechanics but require the principal argument that the most probable configuration of the cHTN (i.e., the highest measurement entropy $H[|\psi\rra]$ of the cHTN) is likely realizable \cite{JHAP1,JHAP2}.

To conclude this article, we qualify the most probable real-time evolution of $|\psi(t)\ra$ in a generic state $|\psi\rra_L$ of Eq. (\ref{eq:rra}) in the Lorentzian regime of the cHTN in the presence or absence of matter beyond the gravity perturbation and consider its physical meanings.
(Here, the same qualification is applicable to the most probable imaginary-time evolution in the Euclidean regime of the cHTN.)

When we extremize the generally covariant Lorentzian action (\ref{eq:H}) of the cHTN with respect to $|\psi\rra_L$, there are two distinct tendencies.
First, the measurement entropy $H[|\psi(t)\ra]$ of the cHTN tends to be maximized at every real-time instance $t$.
Second, with respect to real time, the cHTNs tend to diversify into distinct ones with equal statistical weight: the more non-trivial real-time evolution is, the more entropy of the temporal part of $|\psi\rra_L$ is generated.
In the most probable real-time evolution of $|\psi(t)\ra$, these two distinct tendencies of the measurement entropy $H[|\psi\rra_L]$ stem from the equilibrium Boltzmann principle in the Euclidean regime of the HTN and the non-equilibrium second law in the Lorentzian regime of the HTN, respectively, in a generally covariant manner.
Here, the general covariance of $I_L[|\psi\rra_L]$ with respect to the bulk isometries (i.e., the boundary conformal transformations) is the first law.

Finally, we note that, in general relativity governed by the Einstein field equations, the counterpart of our holographic formulation of the Lorentzian spacetime exists in the real-time evolution of black holes, where the second law of gravity is the area theorem of the event horizons of black holes \cite{Laws1,Laws2}.

\end{document}